\documentclass[fleqn,usenatbib]{mnras}

\usepackage{newtxtext,newtxmath}

\usepackage[T1]{fontenc}
\usepackage{ae,aecompl}

\usepackage{graphicx}	
\usepackage{amsmath}	
\usepackage{amssymb}	
\usepackage{ulem}

\newcommand{\cb}[1]{\textcolor{red}{{#1}}}

\title[HI content of galaxies via cross-correlations]{Determining the HI content of galaxies via intensity mapping cross-correlations}

\author[L.Wolz et al.]{
L. Wolz,$^{1, 2}$\thanks{E-mail: lwolz@unimelb.edu.au}
C. Blake,$^{2, 3}$
and J.S.B. Wyithe$^{1, 2}$
\\
$^{1}$School of Physics, University of Melbourne, Parkville, VIC 3010, Australia\\
$^{2}$ARC Centre of Excellence for All-Sky Astrophysics (CAASTRO)\\
$^3$Centre for Astrophysics \& Supercomputing, Swinburne University of Technology, P.O. Box 218, Hawthorn, VIC 3122, Australia\\
}

\date{Accepted XXX. Received YYY; in original form ZZZ}

\pubyear{2016}

\begin{document}
\label{firstpage}
\pagerange{\pageref{firstpage}--\pageref{lastpage}}
\maketitle

\begin{abstract}
We propose an innovative method for measuring the neutral hydrogen (HI) content of an optically-selected spectroscopic sample of galaxies through cross-correlation with HI intensity mapping measurements.  We show that the HI-galaxy cross-power spectrum contains an additive shot noise term which scales with the average HI brightness temperature of the optically-selected galaxies, allowing constraints to be placed on the average HI mass per galaxy. This approach can estimate the HI content of populations too faint to directly observe through their 21cm emission over a wide range of redshifts.  This cross-correlation, as a function of optical luminosity or colour, can be used to derive HI-scaling relations. We demonstrate that this signal will be detectable by cross-correlating upcoming Australian SKA Pathfinder (ASKAP) observations with existing optically-selected samples.  We also use semi-analytic simulations to verify that the HI mass can be successfully recovered by our technique in the range $M_{\rm HI} > 10^8 M_\odot$, in a manner independent of the underlying power spectrum shape.  We conclude that this method is a powerful tool to study galaxy evolution, which only requires a single intensity mapping dataset to infer complementary HI gas information from existing optical and infra-red observations.
\end{abstract}

\begin{keywords}
galaxies: evolution -- radio lines: galaxies -- large-scale structure of Universe
\end{keywords}



\section{Introduction}

Galaxy evolution is driven by a range of processes including star formation, feedback from supernovae and active galactic nuclei, and environmental effects such as gas stripping and galaxy mergers. Galactic star formation activity is fuelled by cold gas reservoirs and the star formation efficiency is driven by the cooling and cycling mechanism of the gas within the galaxy. Before collapsing into stars, atomic hydrogen (HI) is converted into its molecular phase. Understanding the mechanisms of atomic to molecular gas conversion and their relation to the star formation mechanism is the goal of theorists and observers alike.

In the local Universe ($z<0.2$), the Cosmic HI density and the HI mass function has been determined by blind HI surveys via their 21cm-line emission down to masses $10^7 M_\odot$ (HIPASS \citealt{2005MNRAS.359L..30Z}, ALFALFA \citealt{2010ApJ...723.1359M}). Targeted surveys have combined HI detections of galaxies with their UV and optical counterparts which estimate the star formation rate (SFR) of the objects. The derived HI scaling relations establish strong relations between the atomic gas fractions and their stellar masses and star forming activities, see e.g. \cite{2010MNRAS.403..683C}. Spectral stacking studies in \cite{2015MNRAS.452.2479B} suggest that the SFR is the primary driver of the correlations. It has also been shown that there exists a strong environmental dependence of the HI scalings, e.g. in \cite{2011MNRAS.415.1797C} and \cite{2017MNRAS.466.1275B} which is driven by HI stripping processes caused by the dynamics in dense environments \citep{2015MNRAS.448.1715J}. Molecular hydrogen gas is traced via the CO(1-0) line emission and \cite{2011MNRAS.415...32S,2016MNRAS.462.1749S} demonstrated that the molecular gas fraction is strongly correlated with SFR rather than stellar mass of their hosts.

The evolution of neutral gas with cosmic time is still poorly observationally constrained, given that the highest confirmed HI detection is at $z=0.376$ \citep{2016ApJ...824L...1F}.  Therefore, although it is well-established that the cosmic star formation rate peaks at $z \sim 2$  from the combination of UV and IR observations \citep{2014ARA&A..52..415M}, it is less clear how the evolution of the SFR is linked to the cold gas in either atomic or molecular form.  The HI abundance history is also constrained for redshifts $z < 0.5$ by HI stacking techniques \citep{2013MNRAS.435.2693R, 2013MNRAS.433.1398D, 2016MNRAS.460.2675R} and by HI column density measurements from Damped Lyman-alpha systems at higher redshifts \citep{2006ApJ...636..610R, 2009ApJ...696.1543P, 2011ApJ...732...35M, 2012A&A...547L...1N}.  Although the high-redshift behaviour of the HI density remains poorly constrained, the observations indicate a relatively low redshift evolution which is therefore uncorrelated with the SFR history.  This suggests that the molecular gas phase, and its correlations, are the key to understanding galaxy evolution across Cosmic time.

The latest numerical simulations include refined models of the cold gas evolution and recipes for the hydrogen partitioning process in both semi-analytic (e.g. \citealt{2011MNRAS.418.1649L, 2014MNRAS.442.2398P, 2010MNRAS.409..515F}) and hydrodynamical approaches such as EAGLE \citep{2016MNRAS.456.1115B, 2016arXiv160406803C}. Simulated global HI properties are in good agreement with $z=0$ observations, within their volume and resolution limitations. Better higher-redshift observations can test the predicted redshift evolution of atomic and molecular gas in different scenarios, and distinguish between star formation and feedback recipes. 

The Square Kilometre Array\footnote{http://skatelescope.org/} (SKA), and its pathfinder projects such as the Australian SKA Pathfinder (ASKAP)\footnote{http://www.atnf.csiro.au/projects/askap/index.html}, will provide key insights into HI evolution beyond current redshift limitations within the next decade.  A rich variety of 21cm experiments are proposed such as spectral stacking, HI line absorption, HI galaxy surveys and intensity mapping experiments \citep{Santos:2015vi}. Intensity mapping describes observations of the unresolved flux of redshifted 21cm emission on large scales, converted into 3D diffuse maps of the HI density \citep{2004MNRAS.355.1339B, 2010Natur.466..463C}. The primary goal of intensity mapping experiments is cosmological analysis to measure the baryon acoustic oscillation scales and the large scale structure fluctuations to constrain current cosmological models \citep{Chang:2007xk, Wyithe:2007rq}. The intensity maps depend on both the cosmic HI density as a function of redshift, and the bias of the HI density with respect to the underlying structure.  The HI intensity mapping signal has been detected in cross-correlation with the WiggleZ Dark Energy Survey at $z\approx0.8$ in \cite{2013ApJ...763L..20M} which also provided the first measurement of the HI density at this redshift \citep{2013MNRAS.434L..46S}. Many new intensity mapping experiments are in preparation spanning a wide range of redshifts, such as BINGO \citep{2012arXiv1209.1041B}, CHIME \citep{Bandura2014}, and Tianlai \citep{Chen:2012xu}. The cross-correlation with optical galaxies has the advantage of being independent of the systematic effects of the intensity maps caused by astrophysical signal contaminations, instrument uncertainties, and data processing. Therefore, the measurement of the cross-correlation is highest priority in the upcoming intensity mapping projects.

In this paper, we propose a new interpretation of the cross-correlation signal of intensity maps with galaxy surveys, which uses this signal to measure the average HI content of the optically-selected galaxies. This measurement is not based on the large-scale structure cross-correlation signal but originates from the shot noise term in the power spectrum measurement. The shot noise term is caused by Poisson sampling noise whose amplitude scales as the inverse of the sampled galaxy density. In the case of intensity mapping cross-correlations, the cross-shot noise term is additionally weighted by the average brightness temperature of the galaxy population. This means by measuring the shot noise amplitude on the cross-power spectrum, we can determine the average HI content of the galaxy sample. From one intensity mapping data set, the cross-correlation with multiple galaxy survey catalogues binned into selection criteria, such as colour, size or stellar mass, can supply multiple HI scaling relations. These HI measurements can be obtained for redshifts inaccessible in direct detection by the current generation of telescopes and provide new insights into galaxy evolution mechanisms.

We present a derivation of the HI cross-shot noise term in Sec.~\ref{Sec_theory}, and demonstrate the detectability of the signal with upcoming ASKAP observations in Sec.~\ref{Sec_forecast}.  In Sec.~\ref{Sec_sim} we present semi-analytical simulations for creating mock HI intensity maps and colour-selected galaxy catalogues at $z=0.9$.  We use these mock catalogues in Sec.~\ref{Sec_fitting} to demonstrate that we can successfully recover the HI shot noise contribution to the cross-power spectra as a function of colour, using a model-independent fitting pipeline.  We summarise and conclude in Sec.~\ref{Sec_concl}.
\section{Power Spectrum formalism}
\label{Sec_theory}
In this section we derive the contribution to the cross-power spectrum of an HI intensity mapping survey and an overlapping optical galaxy redshift survey, that arises from the imprint of the HI content of the optically-selected galaxies in the intensity map.  

The HI brightness temperature $T_{\rm HI}(\bmath{x})$ at position $\bmath{x}$, resulting from HI mass density $\rho_{\rm HI}(\bmath{x})$, is given by
\begin{equation}
T_{\rm HI}(\bmath{x}) =  \frac{3 h c^3 A_{10}}{32 \pi m_H k_B \nu_{21}^2} \frac{(1+z)^2}{H(z)} \times \rho_{\rm HI}(\bmath{x})= C \times \rho_{\rm HI}(\bmath{x})
\label{eqhitemp}
\end{equation}
in terms of Planck's constant $h$, the speed of light $c$, the Einstein coefficient $A_{10}$, the mass of the hydrogen atom $m_H$, Boltzmann's constant $k_B$, the rest-frame frequency of the transition $\nu_{21}$, and the Hubble parameter $H(z)$ at redshift $z$ (e.g. \cite{2015ApJ...803...21B}).

We assume that the HI is contained within $N$ galaxies with number density distribution $n_g(\bmath{x})$ across the observed volume $V$.  We divide the volume into $N_c$ cells with positions $\bmath{x}_i$, such that the number of galaxies in the $i^{\rm th}$ cell is $N(\bmath{x}_i) = n_g(\bmath{x}_i) \times (V/N_c)$.  If $M(\bmath{x}_i)$ is the HI mass associated with each galaxy in cell $i$, the HI brightness temperature is given according to Equ.~\ref{eqhitemp} as
\begin{equation}
T_{\rm HI}(\bmath{x}_i) = C \, N(\bmath{x}_i) \, M(\bmath{x}_i) \, (N_c/V)
\label{Equ_TtoM}
\end{equation}

We consider the cross-correlation between the galaxy density contrast
\begin{equation}
\delta_g(\bmath{x}_i) = \frac{N(\bmath{x}_i) - \langle N \rangle}{\langle N \rangle}
\end{equation}
and the brightness over-temperature
\begin{equation}
\delta_T(\bmath{x}_i) = T_{\rm HI}(\bmath{x}_i) - \langle T_{\rm HI} \rangle
\end{equation}
where the angled brackets denote mean values, and $\langle N \rangle = N/N_c$.  Our estimator for the cross-power spectrum $P_\times$ at wavevector $\bmath{k}$ between the gridded galaxy distribution and intensity map is
\begin{equation}
\hat{P}_\times(\bmath{k}) = V \, {\rm Re} \lbrace \tilde{\delta}_g(\bmath{k}) \, \tilde{\delta}_T^*(\bmath{k}) \rbrace
\end{equation}
The Fast Fourier Transforms of $N(\bmath{x}_i)$ and $T_{\rm HI}(\bmath{x}_i)$ are given by
\begin{equation}
\tilde{N}(\bmath{k}) = \sum_i N(\bmath{x}_i) \, e^{i \bmath{k}.\bmath{x}_i}
\end{equation}
\begin{equation}
\tilde{T}(\bmath{k}) = \sum_i T_{\rm HI}(\bmath{x}_i) \, e^{i \bmath{k}.\bmath{x}_i} = \frac{C \, N_c}{V} \sum_i N(\bmath{x}_i) \, M(\bmath{x}_i) \, e^{i \bmath{k}.\bmath{x}_i}
\end{equation}
The expectation value of our cross-power spectrum estimator is then
\begin{eqnarray}
\langle \hat{P}_\times(\bmath{k}) \rangle &=& V \, {\rm Re} \lbrace \langle \tilde{\delta}_g(\bmath{k}) \, \tilde{\delta}_T^*(\bmath{k}) \rangle \rbrace \nonumber \\ 
&=& \frac{C}{N} \sum_{i,j} \langle \left( N_i - \langle N_i \rangle \right) \left( N_j M_j - \langle N_j M_j \rangle \right) \rangle \, e^{i \bmath{k}.(\bmath{x}_i - \bmath{x}_j)} \nonumber \\
&=& \frac{C}{N} \sum_{i,j} \left( \langle N_i N_j M_j \rangle - \langle N_i \rangle \langle N_j M_j \rangle \right) \, e^{i \bmath{k}.(\bmath{x}_i - \bmath{x}_j)}
\end{eqnarray}
where for brevity we have written $N_i \equiv N(\bmath{x}_i)$ and $M_i \equiv M(\bmath{x}_i)$, and omitted the real part.

For the terms with $j=i$, if we assume that the number of galaxies in each cell is uncorrelated with their HI mass, we have $\langle N_i \, M_i \rangle = \langle N_i \rangle \langle M_i \rangle$, which allows us to write assuming Poisson statistics:
\begin{equation}
\langle N_i^2 \, M_i \rangle - \langle N_i \rangle \langle N_i \, M_i \rangle = \langle N_i \, M_i \rangle
\end{equation}
For the terms with $j \ne i$, we define the galaxy-temperature cross-correlation function $\xi_\times$ as
\begin{equation}
\langle N_i \, N_j \, M_j \rangle - \langle N_i \rangle \langle N_j \, M_j \rangle = \langle N_i \rangle \, \xi_\times(\bmath{x}_i, \bmath{x}_j) \end{equation}
where $\xi_\times$ can be related to the cross-power spectrum as
\begin{equation}
\xi_\times(\bmath{x}_i, \bmath{x}_j) = \frac{1}{(2\pi)^3} \int P_\times(\bmath{k}) \, e^{-i \bmath{k}.(\bmath{x}_i - \bmath{x}_j)} \, d^3\bmath{k}
\end{equation}
After some algebra we obtain
\begin{equation}
\langle \hat{P}_\times(\bmath{k}) \rangle = P_\times(\bmath{k}) + \frac{C}{N} \sum_i N(\bmath{x}_i) \, M(\bmath{x}_i)
\end{equation}
such that the estimator depends on both the underlying clustering power spectrum and a shot noise term SN which depends on the HI content of the galaxy sample.  This can be re-expressed in terms of the average brightness temperature $\overline{T}_{{\rm HI},g}$ generated by the optical galaxies:
\begin{equation}
{\rm SN} = \frac{C}{N} \sum_i N(\bmath{x}_i) \, M(\bmath{x}_i) = \frac{\overline{T}_{{\rm HI},g}}{n_g}
\label{SN}
\end{equation}
The above equation demonstrates that the cross-shot noise is a direct measure of the brightness temperature which is connected to the HI masses located in the galaxy sample via Equ.~\ref{Equ_TtoM}.

Our models for the auto- and cross-power spectra of the galaxies and HI brightness temperature are defined as 
\begin{eqnarray}
P_g(k) &=& b_g(k)^2 \, P_m(k) \nonumber \\
P_T(k) &=& \overline{T}_{\rm HI}^2 \, b_{\rm HI}(k)^2 \, P_m(k) \nonumber \\
P_\times(k) &=& b_\times(k) \, \bar b_g \, \overline{T}_{\rm HI} \, \bar b_{\rm HI} \, P_m(k) 
\label{eqpkmod}
\end{eqnarray}
in terms of the galaxy bias $b_g(k)$, HI bias $b_{\rm HI}(k)$, mean HI brightness temperature $\overline{T}_{\rm HI}$, and matter power spectrum $P_m(k)$. The cross-correlation power spectrum is determined by the large-scales amplitudes of the biases $\bar b_{\rm HI}$ and $\bar b_g$, and $b_\times(k)$ is used to express the scale-dependence of the bias.  In terms of this model, $\langle \hat{P}_\times \rangle = P_\times + {\rm SN}$. For intensity mapping observations, the mean HI brightness temperature can be expressed as a function of the cosmic HI energy density $\Omega_{\rm HI}$ as
\begin{equation}
\overline{T}_{\rm HI}= C \times \Omega_{\rm HI} \, \rho_{\rm crit}.
\end{equation}
We note that the combination $\overline{T}_{\rm HI} \, b_{\rm HI}$ is degenerate, and the individual contributions cannot be determined by HI intensity mapping power spectrum measurements. 

\section{Detectability forecast}
\label{Sec_forecast}
We assessed the potential of upcoming surveys to detect the shot noise contribution to the cross-power spectrum by computing a forecast using the planned DINGO survey by ASKAP as an example.  The overlapping optical galaxy dataset could be provided by the Galaxy and Mass Assembly (GAMA) survey \citep{2011MNRAS.413..971D} in the same regions.

The noise power spectrum $P_N$ for intensity mapping by an interferometer is derived by \cite{2006ApJ...653..815M} and \cite{2015ApJ...803...21B}.  We write the relation in the form
\begin{equation}
P_N(k_\perp) = \sigma_T^2(k_\perp) \, V_{\rm pix} \, d^2u / \Omega_{\rm FOV}
\label{eqpnoise}
\end{equation}
where $k_\perp$ is the wavevector component of the power spectrum in the plane of the sky, $d^2u = 2\pi u \, du$ is a 2D pixel in visibility space where $k_\perp = 2\pi u/r$ and $r$ is the co-moving distance to the observed redshift, $\Omega_{\rm FOV}$ is the field-of-view of an observation in steradians, and $V_{\rm pix}$ is the 3D volume corresponding to $\Omega_{\rm FOV}$ and the frequency width of each channel $\Delta \nu$, which we evaluate as
\begin{equation}
V_{\rm pix} = r^2 \, \Omega_{\rm FOV} \, r_\nu \, \frac{\Delta \nu}{\nu}
\end{equation}
where $r_\nu = \frac{c \, (1+z)}{H(z)}$.  The temperature variance $\sigma_T^2$ in Equ. \ref{eqpnoise} is given by
\begin{equation}
\sigma_T^2(k_\perp) = \frac{\lambda^2 \, T_{\rm sys}}{A_{\rm dish} \sqrt{\Delta \nu \, n(u) \, d^2u \, t_{\rm int} \, N_{\rm beam} \, N_{\rm pol}}}
\end{equation}
where $\lambda = 0.21 \times (1+z)$ m is the observed wavelength at redshift $z$, $T_{\rm sys}$ is the system temperature, $A_{\rm dish}$ is the collecting area of each dish, $t_{\rm int}$ is the integration time, the telescope has $N_{\rm beam}$ beams and $N_{\rm pol}$ polarizations, and $n(u)$ is the number density of baselines in visibility space normalized such that
\begin{equation}
\int n(u) \, d^2u = N_{\rm dish} (N_{\rm dish} - 1)/2
\end{equation}
in terms of the total number of dishes $N_{\rm dish}$.  We average the noise power spectrum in spherical $k$-bins by computing
\begin{equation}
P_N(k) = \int P_N(k_\perp) \, d\Omega_k = \int_0^k \frac{k_\perp \, P_N(k_\perp)}{k \, \sqrt{k^2 - k_\perp^2}} \, dk_\perp
\end{equation}

For ASKAP observations we assume a total collecting area $A_{\rm tot} = N_{\rm dish} \, A_{\rm dish} = 4000$ m$^2$ composed of $N_{\rm dish} = 36$ dishes with $N_{\rm beam} = 30$ beams and $N_{\rm pol} = 2$ polarizations covering a field-of-view $\Omega_{\rm FOV} = 30$ deg$^2$, an instrumental system temperature $T_{\rm inst} = 90$ K and frequency channel width $\Delta \nu = 20$ kHz.  We consider two components: a ``deep'' survey mapping total survey area $\Omega_{\rm surv} = 150$ deg$^2$ and redshift range $0 < z < 0.26$ with total observing time $t_{\rm tot} = 2500$ hrs, and an ``ultra-deep'' survey mapping $\Omega_{\rm surv} = 60$ deg$^2$ and $0.1 < z < 0.43$ with $t_{\rm tot} = 5000$ hrs.  We take the effective redshift for the noise calculation as the mean redshift in the range.  We derive the integration time as
\begin{equation}
t_{\rm int} = t_{\rm tot} \times \frac{\Omega_{\rm FOV}}{\Omega_{\rm surv}}
\end{equation}
and obtain the total system temperature after adding in the sky brightness as
\begin{equation}
T_{\rm sys} = T_{\rm inst} + 60 {\rm K} \left( \frac{300 \, {\rm MHz}}{\nu} \right)^{2.55}
\end{equation}
where $\nu = (1400 \, {\rm MHz})/(1+z)$ is the observing frequency.

The noise in the measurement of the cross-power spectrum $P_\times(k)$ of the intensity map and an overlapping optical spectroscopic redshift survey, in a Fourier bin $k$ of width $\Delta k$ containing $N_{\rm mode}$ unique Fourier modes, is
\begin{equation}
\sigma(P_\times) = \frac{1}{\sqrt{2 \, N_{\rm mode}}} \sqrt{P_\times^2 + \left( P_T + P_N \right) \left( P_g + \frac{1}{n_g} \right)}
\end{equation}
in terms of the temperature power spectrum $P_T(k)$ and galaxy power spectrum $P_g(k)$, where $1/n_g$ is the galaxy shot noise power in terms of the galaxy number density $n_g$.  We use Equ.~\ref{eqpkmod} to model these power spectra.  The number of unique modes is
\begin{equation}
N_{\rm mode} = \frac{V_{\rm surv}}{(2\pi)^3} \, 2\pi k^2 \, \Delta k
\end{equation}
in terms of the total survey volume $V_{\rm surv}$, which we evaluate in the survey cone of solid angle $\Omega_{\rm surv}$ using a fiducial flat $\Lambda$CDM cosmological model with matter density $\Omega_{\rm m} = 0.3$.  For our forecast we assume representative values $b_g = 1$, $b_{\rm HI} \Omega_{\rm HI} = 0.43 \times 10^{-3}$ \citep{2013MNRAS.434L..46S}, $b_\times(k)=1$ and $n_g = 10^{-3} \, h^3$ Mpc$^{-3}$.  We compute the shot noise contribution SN using Equ.~\ref{SN} assuming that each optical galaxy contains HI mass $M_{\rm HI} = 10^9 \, h^{-2} \, M_\odot$ and generate the matter power spectrum $P_m(k)$ using a {\it Planck} fiducial cosmological model, and assume:
\begin{equation}
\overline{T}_{\rm HI} = 197 \, {\rm mK} \times \frac{(1+z)^2}{H(z)}
\end{equation}
(e.g. \cite{2015ApJ...803...21B}).

We evaluate the forecast in 30 logarithmically spaced $k$-bins between $k = 0.1$ and $100 \, h$ Mpc$^{-1}$. We note that the interferometer does not provide any measurements for scales $k < k_{\perp, {\rm min}} = 2\pi u_{\rm min}/r$, where $u_{\rm min}$ is the minimum visibility.  We compute an average ASKAP visibility distribution $n(u)$ given the 2D baseline distribution of the array.

The cross-power spectrum signal $P_\times(k)$ and noise $\sigma(P_\times)(k)$ for these configurations is plotted in Fig.~\ref{Fig_bias}.  We see that the ASKAP-DINGO intensity-mapping observations, cross-correlated with an optical survey such as GAMA, provide accurate measurements of the cross-power spectrum and the ability to detect the cross-shot noise signal with high significance for small scale modes $k > 10 \, h$ Mpc$^{-1}$ where the shot noise is dominating the observed signal, such that we do not require to model the underlying clustering power spectra.

\begin{figure}
\centering
\includegraphics[width=0.5\textwidth, clip=true, trim= 0 0 0 0 ]{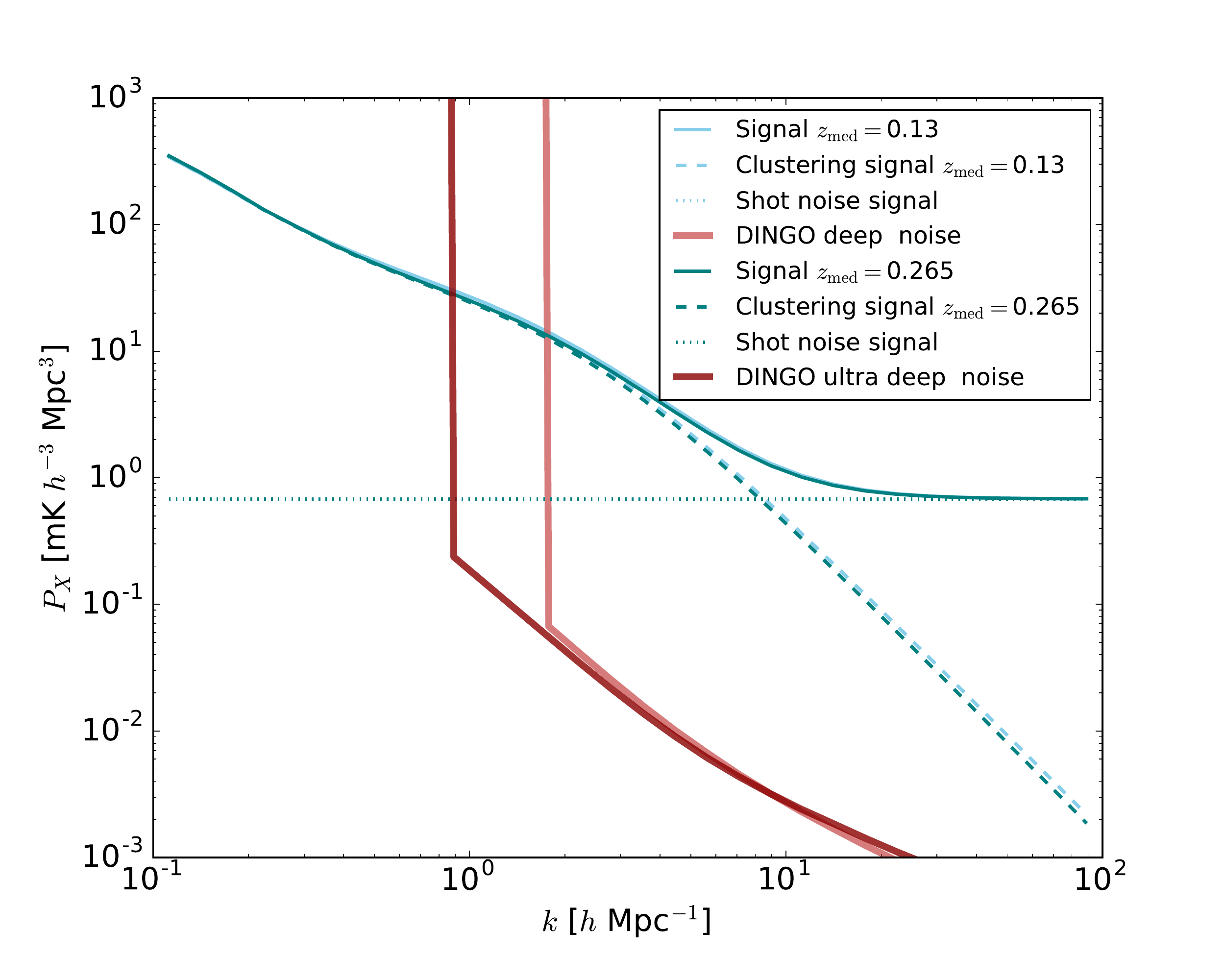}
\caption{The forecast for a intensity mapping cross-power spectrum measurement between the ASKAP-DINGO survey and an overlapping optical survey, for the ``deep'' and ``ultra-deep'' DINGO configurations. The HI cross-shot noise contribution is marked as the dotted line assuming an average HI mass $10^9 \, h^{-2} \, M_\odot$ per optical galaxy.  The noise in the cross-power spectrum is marked as the red solid lines, which are up to three orders of magnitudes smaller than the expected shot noise signal for both DINGO surveys.}
\label{Fig_noise}
\end{figure}

\section{Simulations}
\label{Sec_sim}
For illustration of our detection technique, we employ the simulation presented in \cite{2016MNRAS.458.3399W} which is based on the semi-analytic models used for SAGE, see \cite{2016ApJS..222...22C}. The simulation box is of size $(500 \, {\rm Mpc}/h)^3$ with  particle mass resolution of $8.6 \times 10^{8}h^{-1} M_\odot$. The star-formation model follows the Blitz-Rosolowsky  recipe, see \cite{2006ApJ...650..933B} for details. In summary, the star formation activity depends on the gas surface density, the molecular hydrogen fraction and the molecular gas depletion time.  In this model, the partitioning of the hydrogen in atomic and molecular state\cb{s} is determined by the hydrostatic pressure of the galaxy derived by stellar surface density, gas surface density and the gas velocity dispersion. For a detailed description of the simulations and a study of their galaxy properties and intensity mapping constraints, we refer the reader to \cite{2016MNRAS.458.3399W}.

We showcase the HI measurement concept using a simulation snapshot at $z=0.905$, which is the target epoch of intensity mapping experiments by the GBT, Parkes Telescope and potentially ASKAP. Direct HI measurements at these redshifts are not obtainable with current technologies.  For the purposes of this test, we study the clustering of only central galaxies, removing all satellite galaxies from the simulation for both probes, the HI intensity maps and the optical samples.  The presence of sub-structure complicates the cross-shot noise relations, and we will present a full treatment of these effects in a follow-up study.

We generated the mock datasets as follows:
\begin{itemize}
\item{\textbf{Intensity Maps}}\\
We bin all HI-emitting objects in the simulation box in a grid with resolution $N_{\rm pix}=512^3$ and convert the HI mass in each grid cell into a brightness temperature using Equ.~\ref{eqhitemp}. The resulting intensity maps are in units of $K$.  In order to perform an accurate test of the theory, we neglect telescope effects.
\item{\textbf{Galaxy Samples}}\\
We simulate the galaxy magnitudes in optical and UV filters by applying the photometric mock galaxy code described by \cite{2012ApJ...759...43T} to the semi-analytic simulation output. We cross-correlated mock galaxy samples with the intensity map in bins of $({\rm NUV} -r)$ colour, using the GALEX NUV band and the SDSS optical $r$-band.  The $({\rm NUV} -r)$ colour is a measure of the SFR.  Thus the derived scaling relation can be interpreted as the average HI mass as a function of SFR.  The galaxy sample can alternatively be binned using many other properties such as luminosity, stellar mass, halo mass, environment, molecular content, metallicity or morphology, depending on the available data.  We split the galaxy population by $({\rm NUV}-r)$ colour in bins with divisions at $\{-1, 0, 1, 2, 3, 4, 5, 6\}$, defining samples from blue star-forming to red quiescent galaxies. Our selection process results in galaxy samples with $N_{g}=\{ 774889, 6214418, 694146, 263815, 187110, 662317, 965585\}$ number of objects.
\end{itemize}

\section{HI cross shot noise measurement}
\label{Sec_fitting}
In this section, we present the application of the theoretical concept to the semi-analytic simulation and demonstrate an empirical approach to perform the shot noise measurement from the cross-power spectrum in a manner independent of the underlying cosmological power spectrum by employing the cross-correlation coefficient. 

The cross-correlation shot noise is a scale-independent additive term to the cross-power spectrum. It can be straightforwardly determined from $\hat P_{\times}(k)$ if the amplitude of the shot noise is high enough to dominate the signal and/or the cross-power is measured on small enough scales where the cosmological signal becomes negligible compared to the shot noise. For any other cases, the estimation of the amplitude of the cross-shot noise directly from the cross-power spectrum requires knowledge of the underlying cosmology and bias. 

In this study, we demonstrate the measurement of the shot noise from the observed cross-correlation coefficient $\hat r(k)$ to eliminate the dependence on the underlying clustering signal and, hence, the model dependence. We employ the measured power spectrum of the intensity map, $ P_T(k)$, the galaxy power spectra for the $N$ colour cuts labelled by $i$, $ P_{{\rm g},i}(k)$, and the cross-power spectra of the galaxy samples and intensity map, $\hat P_{\times,i}(k)$ on scales $k < 1.5 \, h$ Mpc$^{-1}$. 

The observed cross-correlation coefficient $\hat r(k)$ includes a shot noise term and can be written as 
\begin{equation}
\hat r_i(k) =\frac{\hat P_{\times,i}(k)}{\sqrt{P_T(k) \, P_{{\rm g},i}(k)}} = \frac{ P_{\times,i}(k) + {\rm SN}_i}{\sqrt{P_T(k) \, P_{{\rm g},i}(k)}} = r_i(k) + \frac{{\rm SN}_i }{\sqrt{P_T(k) \, P_{{\rm g},i}(k)}}
\label{rX}
\end{equation}
Using Equ.~\ref{eqpkmod}, we can rewrite the cross-correlation coefficient $r_i(k)$ as 
\begin{equation}
r_i(k)= \frac {b_{\times,i}(k) \bar b_{g,i} \bar b_{\rm HI}}{ b_{\rm HI}(k) b_{g,i}(k)}
\end{equation}
In order to measure the shot noise contribution from $\hat r_i(k)$, we require a model for the scale-dependence of $r_i(k)$.

If the HI and galaxy bias were both scale-independent, the cross-correlation coefficient could be expected to be only mildly scale-dependent, for instance, caused by environmental effects and to converge to 1 on large scales since the amplitudes of the biases $\bar b_{\rm HI}$ and $\bar b_{\rm gal}$ cancel out. Any excess power above 1 measured in $\hat r_i(k)$ would be due to the shot noise term.

In the case of scale-dependent bias, we need to determine the scale-dependence of the quotient $b_\times(k)/(b_{\rm HI}(k)b_g(k))$. 

For illustrative purposes, let us consider the cross-correlation where the intensity maps have scale-independent bias but the galaxy sample is scale-dependently biased on large scales.  This would, however, result in no upward bias in the cross-correlation of both probes since the increased power of the large scale Fourier modes is not present in the intensity maps and therefore is not correlated. In this case, the scale dependency of the cross-correlation bias would follow $b_\times(k) \sim b_{\rm HI}(k)$ and the shape of the cross-correlation coefficient could be approximated as $r(k)\sim 1/b_{g}(k)$.

In more general terms, the functional form of the bias $b_\times(k)$ of the cross-correlation power spectrum follows the scale-dependence of the less dominantly biased power of both probes. The shape of the resulting cross-correlation coefficient $r(k)$ is driven by the inverse of the bias which is not reflected in the cross-correlation bias, i.e. the inverse of the more dominant scale-dependent power of the bias.

In our simulations, after removal of the satellites, the galaxy bias is fairly scale-independent whereas the HI intensity maps exhibit a strong scale-dependent bias on smaller scales $k>0.5 \, h$ Mpc$^{-1}$ which is caused by the increased clustering of HI-rich objects in dense regions predicted by this star formation recipe. This implies that $b_\times(k)\sim b_g(k)$ and, therefore, the shape of the cross-correlation coefficient $r_i(k)$ is proportional to $1/b_{\rm HI}(k)$. For this case, we can assume that $r_i(k)$ is independent of the galaxy sample $i$ and converges to the same scale-dependent form $\bar r(k) \sim 1/b_{\rm HI}(k)$. In the following, we present an empirical, model independent technique to approximate $\bar r(k)$ using the auto-power spectra.

\begin{figure}
\centering
\includegraphics[width=0.5\textwidth, clip=true, trim= 40 660 660 100 ]{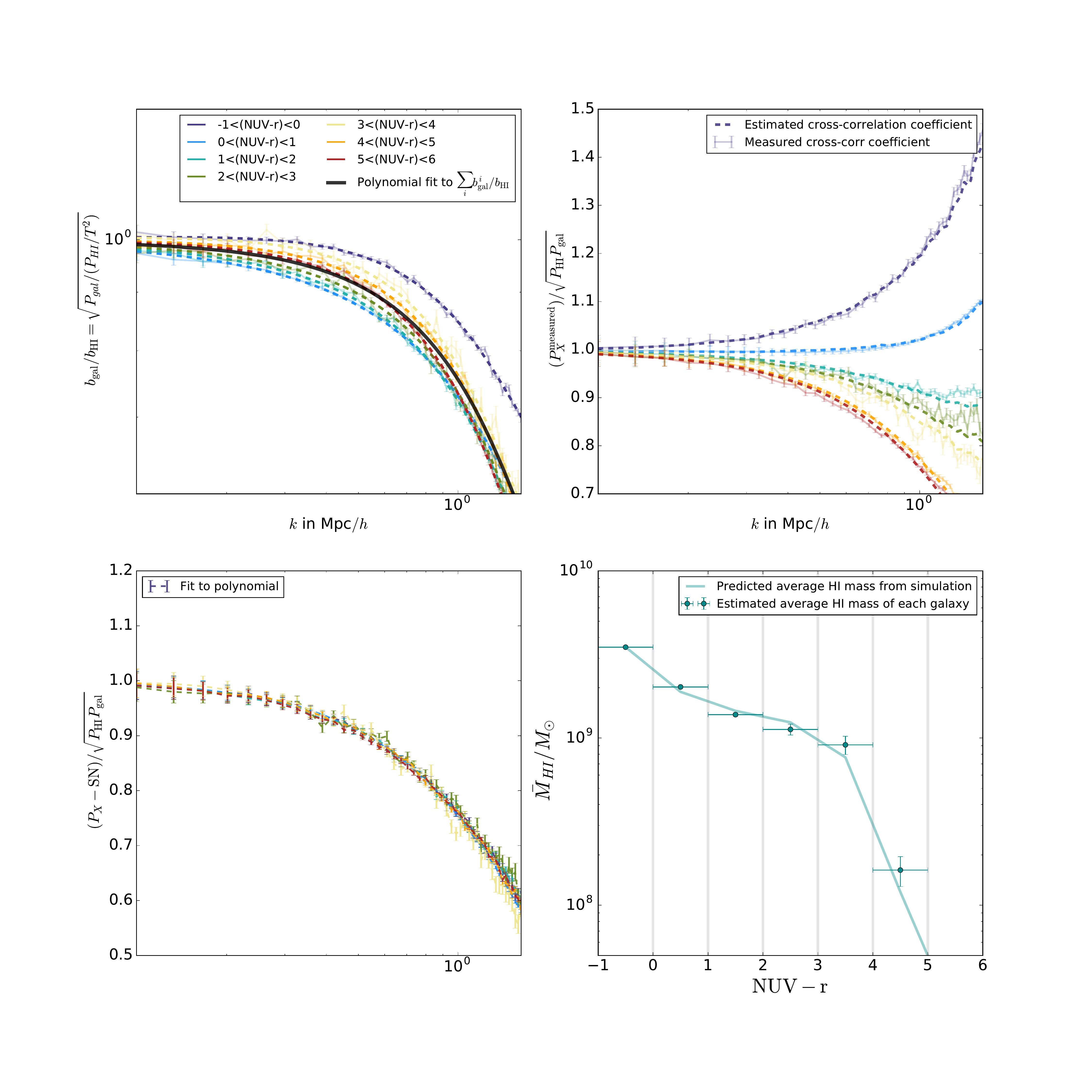}
\caption{We show the quotient of the galaxy bias and the HI bias determined by taking the square-root of the galaxy power spectrum divided by the intensity mapping power spectrum $\sqrt{\hat P_{g,i}(k)/\hat P_T(k)}$. The different colours refer to the power spectra selected by galaxy colour $({\rm NUV}-r)$ which is an indicator for the star-formation rate of the sample, from blue, star-forming to red, quiescent objects. The dashed coloured lines indicate the best-fit polynomial to the quotient of each individual sample and the solid black line marks the best-fit polynomial to the averaged bias quotient.}
\label{Fig_bias}
\end{figure}
\begin{figure}
\centering
\includegraphics[width=0.5\textwidth, clip=true, trim= 640 660 40 100 ]{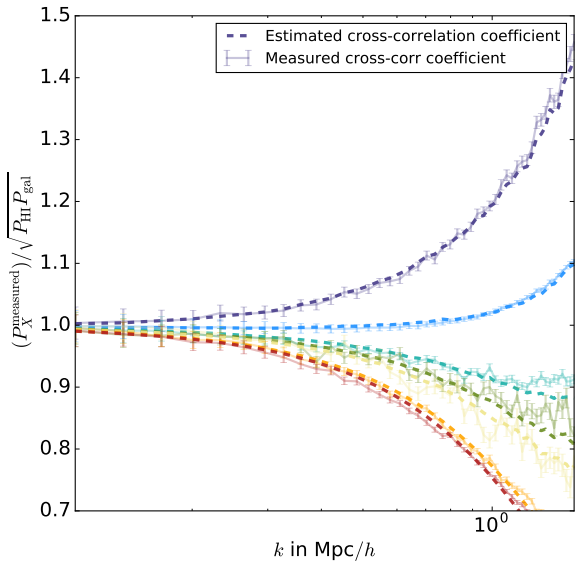}
\caption{The cross-correlation coefficient $\hat r(k)$ of the intensity maps and the different galaxy samples. The colours from blue to red indicate the colour bins in $\Delta({\rm NUV}-r)=1$ from -1 to 6. The splitting of the measurements seen in this plot is caused by the HI cross-shot noise ${\rm SN}=\bar T_i/n_g$. The dashed lines are the the best-fit of the polynomial plus the estimated shot noise parameter $\hat{\rm SN}/\sqrt{P_g(k)P_T(k)}$. }
\label{Fig_crosscorr}
\end{figure}

We estimate the functional scale-dependence of $\bar r(k) \sim 1/b_{\rm HI}(k)$ using  $\sqrt{ P_{g,i}(k)/ P_T(k)}$ which is equal to $B_i(k)=b_{{\rm g}, i}(k) / b_{\rm HI}(k)$ for each population $i$. In Fig.~\ref{Fig_bias}, we show $B_i(k)$ for each colour bin $i$ where each curve has been re-normalised to 1 on large scales. It can be seen that the functional form of the quotient is preserved over the different colour cuts $i$ within a small variance. This confirms that the galaxy bias can be assumed as scale-independent though the amplitude varies considerably with colour cut. We perform a polynomial fit $f(k)=\sum_j^N p_j k^j$ with $N=3$ to the average of all bias quotients $\bar B=\sum_i B_i$ and employ the resulting coefficients ${p_1, p_2, p_3}$ to determine the scale-dependence of the model for $\bar r(k)$. The resulting best fit $f_{\rm bf}(k)$ is shown as the black curve in Fig.~\ref{Fig_bias}. Note that we set $p_0=1$ since an overall normalization factor has already been removed. 

In Fig.~\ref{Fig_crosscorr}, we show the cross-correlation coefficient $\hat r_i(k)$ for all colour cuts $i$ measured as in Equ.~\ref{rX} as the solid lines. We determine the cross-shot noise by fitting the observed $\hat r_i(k)$ to 
\begin{equation}
F_i(k)=f_{\rm bf}(k) + \frac{\hat{\rm SN}_i}{\sqrt{P_{{\rm g},i}(k)P_T(k)}}, 
\label{eqfit}
\end{equation}
where $f_{\rm bf}(k)$ models $\bar r(k)$.
The result of the best fit $F_i(k)$ is plotted in Fig.~\ref{Fig_crosscorr} as the dashed lines. We can see that the best-fit model and measurement are in good agreement. For red galaxy samples with ${\rm NUV}-r>4$, the cross-shot noise amplitude is extremely small since galaxies in these bins are HI-poor which results in a small brightness temperature.

\begin{figure}
\centering
\includegraphics[width=0.5\textwidth, clip=true, trim= 630 100 100 660 ]{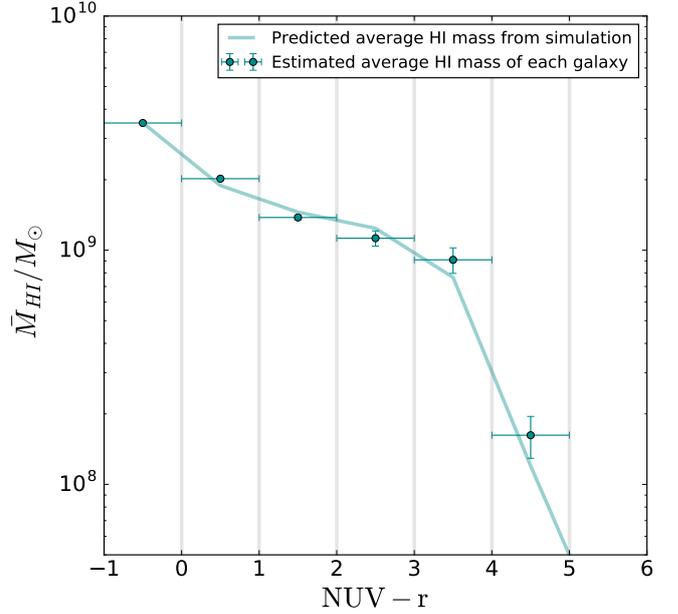}
\caption{The estimated average HI mass per galaxy as a function of the colour $({\rm NUV}-r)$ of the population marked as the circles compared to the simulated model marked with the dashed lines. The errors are given by the model uncertainties of the polynomial fitted to the bias quotient and the variance given by each individual shot noise fit. }
\label{Fig_scaling}
\end{figure}

We can convert the fitted cross-shot noise $\hat{\rm SN}_i$ into an average HI mass per galaxy for each sample $i$ using ${\rm SN}_i= \overline{T}_{{\rm HI, g},i}/n_{{\rm g},i}$ in combination with Equ.~\ref{Equ_TtoM}. The results are shown in Fig.~\ref{Fig_scaling}. The fitted average HI mass $\overline{M}_{{\rm HI},i}$ are marked in symbols and the true values measured directly from the simulation box are given as the solid line. The error bars on this measurement are composed of the standard error given by the weighted least-square fit of the measurement to the model, in addition to the error on the model itself which can be estimated from the covariance of the polynomial fit. The  method performs very well over two orders of magnitude for HI masses above $10^8 M_\odot$. Within the current pipeline, it is not possible to extend HI mass measurements beyond this threshold and for the reddest colour bin, we can not determine an upper limit on the HI mass. This is due to the fact that the HI cross-shot noise lies within the variance of the polynomial fit itself.

\section{Conclusions}
\label{Sec_concl}
In this study, we verify the possibility of determining HI scaling relations for high redshifts using HI intensity mapping cross-correlations with galaxy populations. We show that the shot noise term of the cross-correlation power spectrum is proportional to the average HI mass of the galaxy population. We forecast the detectability of the HI cross-shot noise term of the cross-power spectrum for up-coming ASKAP DINGO observations. We use semi-analytic simulations to demonstrate the measurement of the HI cross-shot noise for a range of different galaxy samples. Our model-independent fitting pipeline can detect the shot noise on scales  $k < 1.5 \, h$ Mpc$^{-1}$ for HI masses above $10^8M_\odot$. Shot noise measurements for $k > 1.5 \, h$ Mpc$^{-1}$ are less sensitive to the fitting method since the noise amplitude dominates the cosmological signal on smaller scales.

In future work, we plan to extend our studies to satellite galaxies, investigating the effect of the central-satellite term on the HI cross-shot noise, within a full halo occupation formalism. We will also develop a more sophisticated method to estimate the scale-dependence of the cross-correlation coefficient by accurate modelling of the HI and galaxy bias terms from their respective auto-power spectrum. 

We conclude that the HI measurements via intensity mapping cross-correlations have a strong potential to deliver a variety of new constraints for galaxy evolution models at high redshifts. The HI shot noise measurements provide a complimentary use of the cross-correlation power spectrum to cosmological tests, independent of the systematic challenges the first generation of intensity mapping experiments are facing.

\section*{Acknowledgements}
We thank Paul Geil and Martin Meyer for useful discussions throughout this project. LW is supported by an ARC Discovery Early Career Researcher Award (DE170100356). This research was conducted by the Australian Research Council Centre of Excellence for All-sky Astrophysics (CAASTRO), through project number CE110001020. 

\bsp	
\label{lastpage}
\bibliographystyle{mn2e}
\bibliography{bib2}
\end{document}